\DeclareUnicodeCharacter{0301}{\'{e}}
\documentclass{article}
\usepackage{tabularx}
\usepackage{spconf,amsmath,graphicx}

\usepackage{threeparttable}
\usepackage{spconf,amsmath,graphicx}
\usepackage{float}
\usepackage{subfigure}
\usepackage{graphicx}
\usepackage{multirow}
\usepackage{booktabs}
\usepackage{tabularx}
\usepackage{hyperref}
\usepackage{amssymb}

\title{Spatial-DCCRN: DCCRN Equipped with Frame-level Angle Feature and Hybrid Filtering for Multi-channel Speech Enhancement}
\name{Shubo Lv$^{1,2}$, Yihui Fu$^1$, Yukai Jv$^{1,2}$, Lei Xie$^1$$^{*}$\thanks{*: corresponding author.}, Weixin Zhu$^2$, Wei Rao$^2$, Yannan Wang$^2$}

\address{
  $^1$Audio, Speech and Language Processing Group (ASLP@NPU), \\
  Northwestern Polytechnical University, Xi'an, China \\
  $^2$Tencent Ethereal Audio Lab, Tencent Corporation, Shenzhen, China}

\copyrightnotice{978-1-6654-7189-3/22/\$31.00~\copyright2023 IEEE}
\begin{document}
%
\maketitle
\begin{abstract}
Recently, multi-channel speech enhancement has drawn much interest due to the use of spatial information to distinguish target speech from interfering signal. To make full use of spatial information and neural network based masking estimation, we propose a multi-channel denoising neural network -- Spatial DCCRN. Firstly, we extend S-DCCRN to multi-channel scenario, aiming at performing cascaded \emph{sub-channel} and \emph{full-channel} processing strategy, which can model different channels separately. Moreover, instead of only adopting multi-channel spectrum or concatenating first-channel's magnitude and IPD as the model's inputs, we apply an \emph{angle feature extraction} module (AFE) to extract frame-level angle feature embeddings, which can help the model to apparently perceive spatial information. Finally, since the phenomenon of residual noise will be more serious when the noise and speech exist in the same time frequency (TF) bin, we particularly design a \emph{masking and mapping filtering} method to substitute the traditional filter-and-sum operation, with the purpose of cascading coarsely denoising, dereverberation and residual noise suppression. The proposed model, Spatial-DCCRN, has surpassed EaBNet, FasNet as well as several competitive models on the L3DAS22 Challenge dataset. Not only the 3D scenario, Spatial-DCCRN outperforms state-of-the-art (SOTA) model MIMO-UNet by a large margin in multiple evaluation metrics on the multi-channel ConferencingSpeech2021 Challenge dataset. Ablation studies also demonstrate the effectiveness of different contributions.
\end{abstract}

\begin{keywords}
multi-channel, Spatial-DCCRN,  speech enhancement
\end{keywords}
\section{Introduction}
\label{sec:intro}

Recently, with the tremendous success of deep learning, speech enhancement has been formulated as a supervised learning problem~\cite{wang2018supervised, weninger2015speech}. 
Meanwhile, multi-channel speech enhancement is gaining increasingly interest due to the utilization of spatial information to distinguish target speech from interfering signal~\cite{zhang2020end, gu2019neural, wang2018combining, fu2021desnet}. Some challenges such as the L3DAS22 challenge~\cite{guizzo2022l3das22} and ConferencingSpeech2021 challenge~\cite{rao2021interspeech} have been recently organized to promote research on multi-channel speech processing. 

A typical strategy is to combine DNNs with traditional beamforming techniques. 
Specifically, TF-mask can be predicted by DNNs, which is used to determine minimum variance distortionless response (MVDR)~\cite{xiao2016study} and generalized eigenvalue (GEV)~\cite{heymann2015blstm} beamforming weights. 
However, as the second stage is purely based on statistical theory and is usually irrelevant to the mask estimation, the pre-estimation error may heavily hamper the subsequent beamforming results~\cite{li2021embedding}. 
More recently,
neural beamformer method, including FasNet~\cite{luo2019fasnet}, EabNet~\cite{li2021embedding} and MIMO-Unet~\cite{ren2021causal}, have shown outstanding performance. EabNet, MIMO-Unet and FasNet employ DNNs to estimate beamforming filters and apply filter-and-sum operation to estimate a single-channel enhanced complex spectrum or waveform.

In frequency domain, for EaBNet~\cite{li2021embedding}, two core modules are designed accordingly, namely EM and BM. The embedding module (EM) learns the 3D spectral and spatial embedding tensor, while the beamforming module (BM) estimates the beamforming weights to implement filter-and-sum operation. Moreover, MIMO-Unet~\cite{ren2021causal} uses a convolutional U-Net to estimate beamforming filters and applies filter-and-sum operation to estimate a single-channel enhanced complex spectrum. Working in the time domain, FasNet~\cite{luo2019fasnet} estimates linear spatial filters for filter-and-sum beamforming.

The spatial information is of vital importance for the multi-channel scenario. However, without any explicite spatial features as input, the above approaches only adopt the multi-channel spectrum as the input of the model and let the network learn spatial information implicitly. Some other strategies adopt reference channel's magnitude and the inter-channel phase difference (IPD) as the input of the network~\cite{gu2019end}. However, only using phase information to reflect channel correlation will lead to apparent information lost. Furthermore, the above methods mostly estimate a group of filters and apply filter-and-sum operation. Yet, when the noise and speech exist in the same TF-bin, the phenomenon of residual noise will be more serious as well.

In this paper, to address the above problems, we propose Spatial-DCCRN for multi-channel speech enhancement. 
The contribution of this work is three-fold, summarized as follows.
\begin{itemize}
    \item We extend the cascaded sub-band and full-band processing strategy of Super wide band DCCRN (S-DCCRN)~\cite{lv2021s} to multi-channel scenario to execute sub-channel and full-channel processing, aiming at benefiting from both local and global channel information processing. Different from the oracle S-DCCRN, we use LSTM in sub/full-channel DCCRN to accept the concatenation of angle feature embedding and encoder's output simultaneously. In addition, the complex mask is replaced with masking and mapping filter (MMF), aiming at denoising and dereverberation simultaneously.
    Compared with monaural processing, sub/full-channel processing strategy in multi-channel scenario can take advantage of the spatial information meticulously and model different channels separately. 
    \item  We design an angle feature extraction (AFE) module, which only adopts the cosIPDs feature as input, aiming at extracting frame-level angle feature. With the help of convolution layers and denseblock, the channel- and time-correlation information can be effectively modeled by the AFE module, which helps the network perceive spatial information apparently.
    \item We design a masking and mapping filtering (MMF) method to replace the typical filter-and-sum operation. Using such cascading strategy, the masking operation aims at dereverberation and coarsely denoising while the mapping operation is designed to further remove residual noise.
    In detail, the masking operation aims at dereverberation and coarsely denoising while mapping operation is designed to remove residual noise. Specifically, we apply sub/full-channel DCCRN to estimate the mapping filters. 
    After that the masking filters are estimated by a group of conv3d blocks which receive the stack of noisy spectrum and mapping filters as the inputs, due to the mapping filters contain the masking filters. With the assistance of conv3d, the magnitude mask of the target channel can be estimated by the input of the MMF module. After applying masking filters to magnitude, the mapping filters are employed on the coarse real / imaginary part to acquire the enhanced speech.
    \end{itemize}
After combining the contributions above, Spatial-DCCRN has surpassed several SOTA models, including FasNet~\cite{luo2019fasnet} and EaBNet~\cite{li2021embedding}, and obtains 0.956 metric score which are composed of STOI and WER and ranks forth on the L3DAS22 challenge dataset. Moreover, 
with the experiments on ConferencingSpeech2021 challenge dataset, our system outperforms baseline by a large margin and also surpasses MIMO-Unet~\cite{ren2021causal} which ranked first in that challenge.

\section{Proposed System}
\label{sec:format}
\subsection{Signal Model}

Assuming ${x^{p}(t)}$, with ${p = 0,... , P-1}$, denotes the time domain noisy and reverberant speech signal at the ${p}$th microphone. The signal model of multi-channel speech enhancement in the short-time Fourier transform (STFT) domain can be given by:
    \begin{equation}\label{form:signal}
    \begin{aligned}
    X_{f,t}=c_{f}S_{f,t}+N_{f,t}
    \end{aligned}
    \end{equation}  
where \{${X_{f,t},S_{f,t},N_{f,t}}$\} denote noisy signal, clean signal and noise respectively with frequency index of ${f\in {1,...,F}}$ and time index of ${t\in {1,...,T}}$. Furthermore, ${c_{f} \in\mathbb{C}^{P\times 1}}$ denotes the relative transfer function (RTF) of the source speech.

\subsection{Multi-Channel DCCRN}
\begin{figure}[t]
    \centering
    \includegraphics[width=1\linewidth]{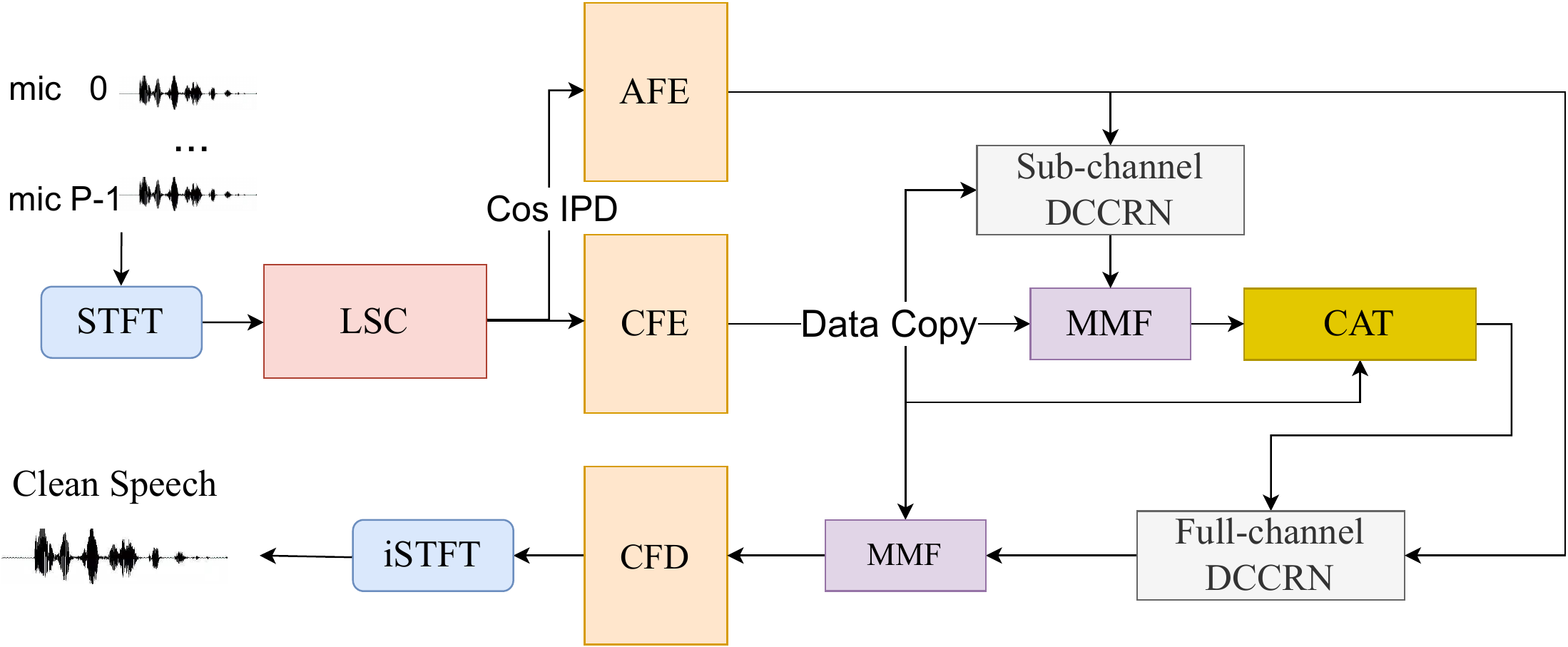}
    \caption{Network structure of the proposed Spatial-DCCRN. "LSC" denotes learnable spectrum compression, "AFE" denotes angle feature extraction, "CFE" denotes complex feature encoder, "CFD" denotes complex feature decoder, "MMF" denotes masking and mapping filtering and "CAT" denotes concatenate.} 
    
    \label{fig:net}
    \end{figure}

Previously, S-DCCRN~\cite{lv2021s} was proposed for super wide-band speech enhancement. The S-DCCRN is equipped with a cascaded sub-band and full-band (SAF) processing module, aiming at benefiting from both local and global frequency information processing. In detail, the SAF module consists of cascaded sub/full-band DCCRN, which substitutes the complex convolution of original DCCRN with group complex convolution. In this paper, inspired by the idea of local and global modeling, we extend S-DCCRN to multi-channel scenario by applying sub-channel and full-channel processing to fully utilize the spatial information. The overall architecture of the proposed Spatial-DCCRN is shown in Fig.~\ref{fig:net}. Similar to the oracle S-DCCRN, the learnable spectrum compression (LSC) module, which consists of a group of trainable compression ratios, is applied to adjust the energy of different frequency bands. The motivation of this design is due to the fact that higher frequency band of the far-field audio is likely to have low energy components. Moreover, the complex feature encoder/decoder module (CFE/CFD) is adopted to extract information from multi-channel complex spectrum. With the help of convolution layers and denseblock~\cite{pandey2020densely}, the CFE/CFD block can refine channel-correlation and time-correlation information. More details of LSC and CFE/CFD can be found in~\cite{lv2021s}. 

Different from the original S-DCCRN, we substitute the complex masking strategy with our proposed masking and mapping filtering (MMF) method, aiming at denoising and dereverberation simultaneously as reverberation is a key issue for far-field speech. Furthermore, we design a learnable angle feature extraction (AFE) module to extract frame-level angle features. Then a modified LSTM layer of sub/full-band DCCRN is applied to accept the concatenation of angle features embedding and encoder's output as input for joint modeling. With the help of the AFE module, the model can model the spatial information apparently. The AFE and MMF modules in our Spatial-DCCRN are introduced in detail in the following.

\subsection{Angle Feature Extraction}

Angle feature is critical for multi-channel enhancement. A widespread strategy is to adpot cosIPDs and reference channel's magnitude as the input's features. However, concatenating those features may increase the difficulty of modeling -- the neural network has to model angle features and magnitude features simultaneously. Another method is to employ a module to estimate the direction of arrival (DOA) of the source speech and noise. However, due to the complex acoustic conditions in real scenario, it is extremely difficult to accurately estimate such DOA features. To this end, we propose a block which only adopts the cosIPDs features as the input to estimate frame-level angle features. 

\begin{figure}[t]
    \centering
    \includegraphics[width=1\linewidth]{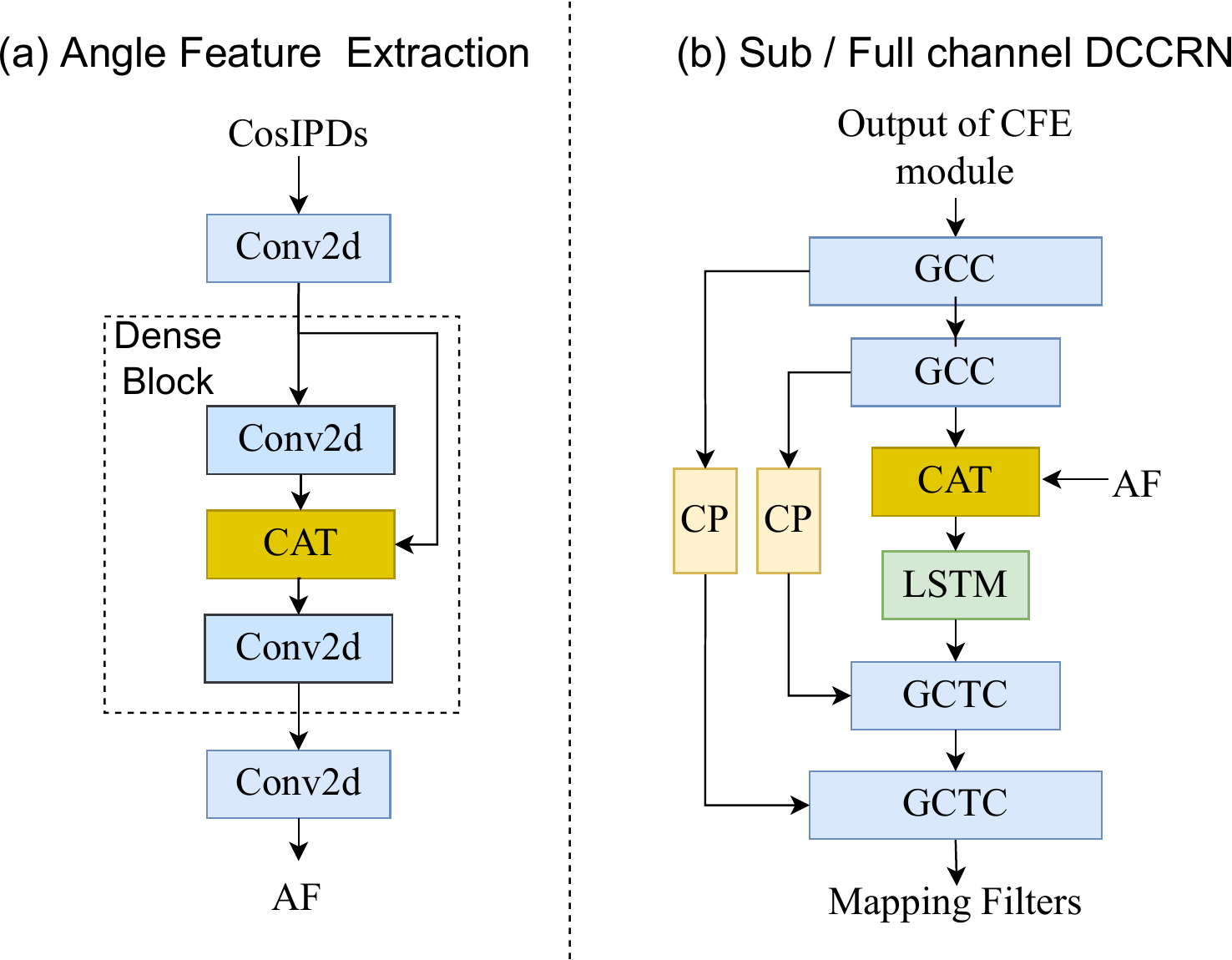}
    \caption{Network structure of angle feature extraction and sub/full channel DCCRN. `GCC` denotes group complex convolution, `GCTC` denotes group complex transpose convolution and `CP` denotes convolution pathway -- a convolution layer among the encoder and decoder~\cite{lv2021dccrn+}. "AF" denotes angle feature embedding.} 
    \label{fig:afe}
    \end{figure}

As shown in Fig~\ref{fig:afe}, the angle feature extraction module is similar to CFE. We employ conv2d with a kernel size of 1 to extract high-dimensional information. Then a dilated dense block
whose depth is 2 is used to capture long-term contextual angle features from time scale. Finally, a conv2d is adopted to extract local angle features. LayerNorm 
and PReLU activation 
are placed after each convolution layer. Afterwards, the LSTM layer of sub/full-channel DCCRN receives the frame-level time-variant angle feature embedding as the input for temporal dependency modeling. With the help of the angle feature extraction module, the complex spectrum enhancement module can apparently perceive the angle information of every frame.

    \begin{figure}[t]
    \centering
    \includegraphics[width=1\linewidth]{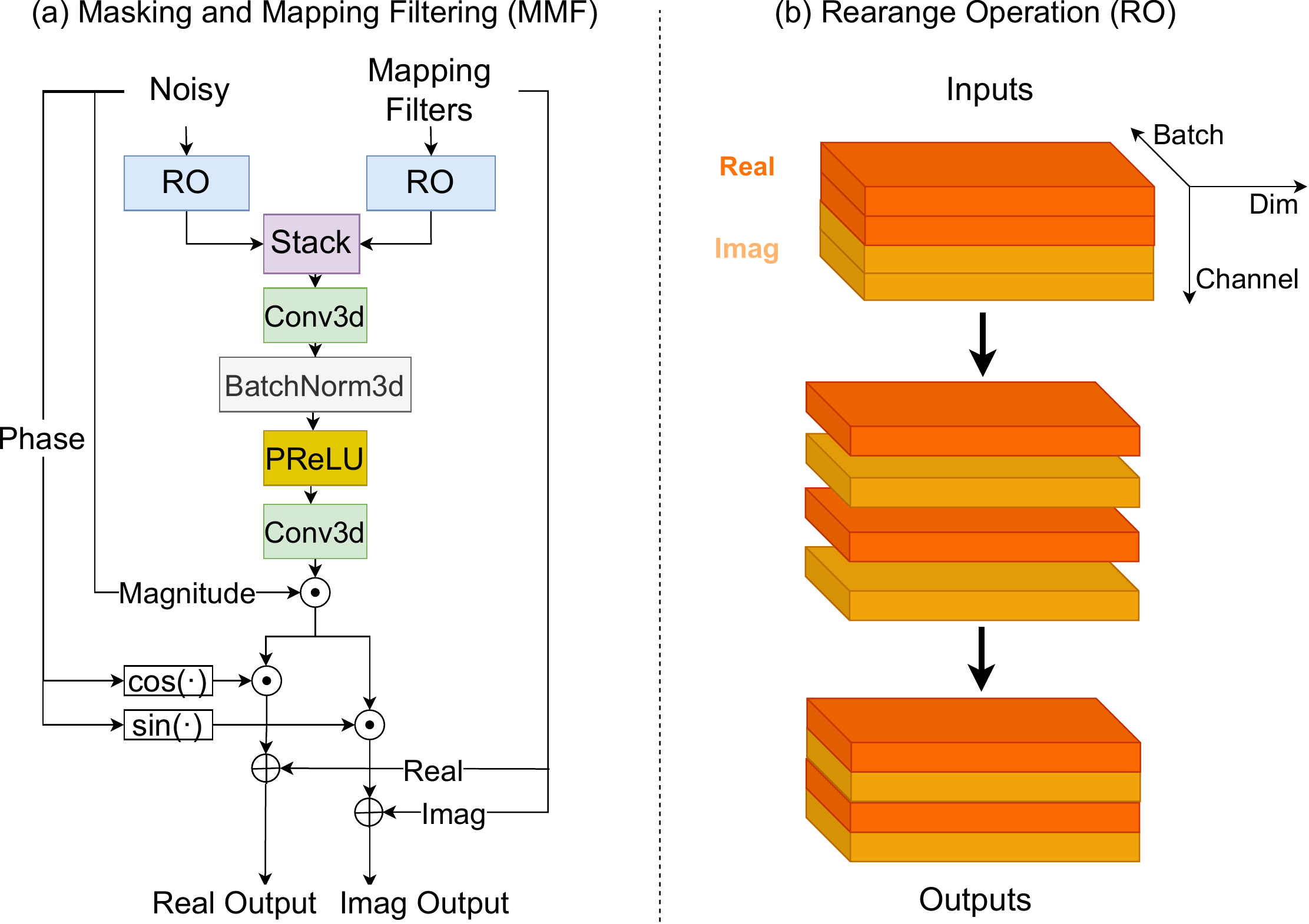}
    \caption{Network structure of masking and mapping filtering.}
    \label{fig:mamf}
    \end{figure}
\subsection{Masking and Mapping Filtering}
Previous studies usually estimate a group of filter weights to do mask based beamforming in frequency or time domain. However, when the noise and speech exist in same TF-bin, residual noise components are relatively hard to be removed. Aiming to suppress the noise and dereverberation simultaneously, we design a novel masking and mapping filtering method. Specifically, we first define the estimated clean speech as
    \begin{equation}\label{form:clean}
    \begin{aligned}
    S_{f,t}={M}'_{f,t}X_{f,t}+M''_{f,t}
    \end{aligned}
    \end{equation}  
where $M'$ and $M''$ denote the masking metrix and mapping metrix respectively. Combining Eq.~(\ref{form:signal}) and Eq.~(\ref{form:clean}), ${M'}$ and ${M''}$ can be described as

    \begin{equation}\label{form:mmf}
    \begin{aligned}
        M'_{f,t} &= 1/c_{f,t} \\
        M''_{f,t} &= -N_{f,t}/c_{f,t}
    \end{aligned}
    \end{equation}  

As a result, the masking operation aims to dereverb while the mapping operation works for removing residual noise. The detailed design of MMF is shown in Figure~\ref{fig:mamf}. Inspired by the collaborative reconstruction module (CRM) in~\cite{li2022glance}, we first employ masking on the magnitude domain and then apply mapping residual operation on the complex domain. The difference between CRM and MMF is that we apply the priority information ($M''$) to estimate $M'$.

As shown in Eq.~(\ref{form:mmf}), ${M''}$ consists of noise and the RTF of speech, which should be estimated by a complex neural network. 
Therefore, the sub/full-channel processing block is employed to 
estimate $M''$. Due to ${M''}$ involves ${M'}$, we stack ${M''}$ with the noisy complex spectrum to estimated ${M'}$. Specifically, we first rearange the input channel to ensure that the real and imaginary parts of the input are alternately placed along channel axis. Then we stack the noisy spectrum and the output of SAF block as the input of the conv3d layer. On the one hand, with the conv3d, the spectral information exists in noisy and ${M''}$ can be well grasped. For another, we set the stride of the dim of mic channel to 2, which can maintain that magnitude mask of the target channel is estimated by its corresponding real and imaginary components. After generating ${M'}$, we apply it to noisy magnitude. Finally, ${M''}$ is applied 
to the coarse real/imaginary part to remove the residual noise.

\subsection{Loss Function}
For the learning objective, we adopt a hybrid loss function strategy. Specifically, besides taking SI-SNR~\cite{luo2019conv} as the time-domain loss function, we also adopt the PHASEN loss to emphasize the phase of T-F bins with higher amplitude, which can help the network to focus on the high amplitude T-F bins where most speech signals are located~\cite{yin2020phasen}.

Finally, the STOI loss~\cite{taal2010short} is applied to directly improve the objective results~\cite{pariente2020asteroid}. Finally, the three losses are optimized jointly by
\begin{equation}
     \setlength{\arraycolsep}{0.3pt}
     \begin{cases}
     \mathcal{L}_{\textbf{PHASEN}} &= \left | |S|^{p} - |\hat{S}|^{p}\right |^{2} + \left | |S|^{p}e^{j\varphi (S)} - |\hat{S}|^{p}e^{j\varphi (\hat{S})}\right |^{2} \\
     \mathcal{L} &= \mathcal{L}_{\textbf{SI-SNR}} + \mathcal{L}_{\textbf{STOI}} + \mathcal{L}_{\textbf{PHASEN}} \\
     \end{cases}
\end{equation}
where ${\hat{S}}$ and ${S}$ denote the network output and clean spectrum respectively. Hyper-parameter ${p}$ is a spectral compression factor empirically set to 0.3. Operator ${\varphi }$ calculates the argument of a complex number.
    \begin{table*}[htb]
    \begin{center} 
    \centering
    
    \normalsize
    \setlength\tabcolsep{12pt}
    \caption{Results of various models and ablation experiments of the proposed model on L3DAS22 dataset. `++MMF` denotes applying AFE and MMF, `Cau` denotes causal, `Ch.` denotes the channel number of inputs, `SR` denotes SI-SNR, `PN` denotes PHASEN and `ST` denotes STOI.}
    
    \vspace{0.5cm}
    \begin{tabular}{lllllllll}
     \toprule 
    \# & Model                                & Cau.    & Ch. & Loss Function             & Para.(M) & STOI  & WER   & Metric \\
    \midrule 
    
    1 & Spatial-DCCRN$^*$                               &  \checkmark     & 4              & SR+PN+ST & 2.34  & 0.913 & 0.086 & 0.914  \\
    2 & ~ + MMF                        &  \checkmark     & 4              & SR+PN+ST & 2.34  & 0.921 & 0.08  & 0.921  \\
    3 & ~ + AFE              &  \checkmark     & 4              & SR+PN+ST & 2.61  & 0.926 & 0.075 & 0.925  \\
    4 & ~ + + MMF &  \checkmark     & 4              & SR+PN+ST & 2.61  & 0.931 & 0.071 & 0.930  \\
    \midrule
    5 & ~ + + MMF &  \checkmark     & 8              & ST              & 2.61  &0.890  & 0.671 & 0.609  \\
    6 & ~ + + MMF &  \checkmark     & 8              & SR             & 2.61  & 0.837  & 0.270 & 0.783     \\
    7 & ~ + + MMF &  \checkmark     & 8              & PN            & 2.61  & 0.941 & 0.057 & 0.941  \\
    8 & ~ + + MMF &  \checkmark     & 8              & PN+ST       & 2.61  & 0.941 & \textbf{0.053} & 0.944  \\
    9 & ~ + + MMF &  \checkmark     & 8              & PN+SR      & 2.61  & 0.946 & 0.056 & 0.945  \\
    10 & ~ + + MMF &  \checkmark     & 8              & SR+ST        & 2.61  & 0.916 & 0.118 & 0.899  \\
    11 & ~ + + MMF &  \checkmark     & 8              & SR+PN+ST & 2.61  & \textbf{0.946} & 0.055 & \textbf{0.946}  \\
    12 & ~ + + MMF                             & $\times$ & 8              & SR+PN+ST & 8.86  & \textbf{0.957} & \textbf{0.045} & \textbf{0.956}  \\
    \midrule
    13 & MIMO-UNet~\cite{ren2021neural}                & $\times$ & 4              & L1                & 5.52  & 0.889 & 0.175 & 0.857  \\
    14 & EaBNet~\cite{li2021embedding}                              &  \checkmark     & 4              & SR+PN+ST & 2.84  & 0.877 & 0.110 & 0.884  \\
    15 & FasNet~\cite{luo2019fasnet}                               &  \checkmark     & 4              & SR+PN+ST & 3.7    & 0.832 & 0.241 & 0.795  \\
     \bottomrule
    \label{tab:l3d}
    \vspace{-1cm}
    \end{tabular} 
    
    \end{center}
    \end{table*}

\section{Experiments}
\label{sec:pagestyle}
\subsection{Datasets}
We first conduct experiments to prove the effectiveness of each proposed sub-modules on the L3DAS22 challenge dataset~\cite{guizzo2022l3das22}. 
The objective of 3D speech enhancement task in this challenge is to enhance speech signals immersed in the spatial sound field of a reverberant office environment collected by 1st order Ambisonics microphone.
Specifically, the dataset contains more than 40000 virtual 3D audio environments with a duration up to 12 seconds each, reaching a total duration of more than 80 hours. Clean utterances are selected from the clean subset of Librispeech~\cite{panayotov2015librispeech} (approximately 53$\%$ male and 47$\%$female speech) 
while the monophonic noise signals come from FSD50K~\cite{fonseca2022fsd50k}. There are a total of 1440 noise sound files, which include 14 transient noise classes and 4 continuous noise classes. The 3D audio signals are generated by convolving the monophonic audio signals with RIRs which is obtained by performing a circular convolution between the recorded sound and the time-inverted analytic signal~\cite{farina2000simultaneous}. Totally, the fixed training and validation sets, with SNR ranging from 6 to 16 db, contain 37,398 utterances (81 h) and 2362 utterances (4 h), respectively.

Meanwhile, the proposed Spatial-DCCRN is trained and evaluated on the ConferencingSpeech 2021 challenge dataset~\cite{rao2021interspeech} to show its robustness on video conferencing multi-talker scenario. 
In the dataset, the source speech data comes from AISHELL-1~\cite{bu2017aishell}, AISHELL-3~\cite{shi2020aishell}, VCTK~\cite{yamagishi2019cstr} and Librispeech(train-clean-360)~\cite{panayotov2015librispeech}. The speech utterances with SNR larger than 15dB are selected for training. The total duration of clean training speech is around 550 hours. The noise dataset is composed of MUSAN~\cite{snyder2015musan} and Audioset~\cite{gemmeke2017audio}, with total duration of about 120 hours. Besides these two open source datasets, 98 real meeting room noise files recorded by high fidelity devices are also used. The training data are generated on-the-fly and segmented into 8 s chunks in one batch with SNR ranging from -5 to 25 dB. 


\subsection{Training setup and baselines}
\label{sec:typestyle}

For the proposed models, the window length, frame shift and future frame are 25 ms, 6.25 ms and 6.25 ms, respectively, resulting in a 37.5 ms processing time. The STFT length is 512. For the 3D speech enhancement task on L3DAS22, all models are trained for 40 epochs with totally 3200h training data. For the models trained on ConferencingSpeech, the total data 'seen' by the model is more than 9900 h after 18 epochs of training. The initial learning rate of all models is 0.001 and will get halved if there is no loss decrease on the validation set.
We also compare the proposed Spatial-DCCRN and its ablation components with other SOTA models on the L3DAS22 dataset. 

The configuration of our Spatial-DCCRN is described as follows. The number of channels for the sub-channel DCCRN is \{32,64,64,64,128,128\}, and the convolution kernel size and stride are set to (5,2) and (2,1) respectively. In addition, the configuration of the full-channel DCCRN is similar to the sub-channel DCCRN except that the channel number of the first layer is 64. One LSTM layer which consists 256 nodes and follows by a 256 $\times$ 256 fully connected layer is adopted to process the concatenation of the encoder outputs of the sub/full-channel DCCRN and the angle feature embeddings. Each encoder/decoder module handles the current frame and one previous frame. The output number of channel of the complex feature encoder/decoder module is 32, and the depth of denseblock is 5. LayerNorm and PReLU are performed after each convolution, except for the last layer of the CFD module. For the AFE module, the number of hidden channels is 16 and the depth of denseblock is 2. For the MMF module, the kernel size, the number of output channels and the step size of the first conv3d are set to (2, 5, 3), 8 and (2, 1, 1) respectively. In addition, the kernel size, output channel and step size of the second conv3d are (1, 5, 3), 1 and (1, 1, 1) respectively. 

\subsection{Experimental results and discussion}

\label{sec:pagestyle}
As presented in Table~\ref{tab:l3d}, ablation studies are conducted to evaluate the effectiveness of different model components of Spatial-DCCRN, including a) Spatial-DCCRN without AFE and MMF (Spatial-DCCRN$^*$), b) Spatial-DCCRN without AFE, c) Spatial-DCCRN without MMF, d) different input channels of the observed signal, e) the non-causal version of Spatial-DCCRN and f) the ablation on different loss function.
For the non-causal model, we substitute the LSTM in Spatial-DCCRN with BLSTM and look ahead one frame in each convolution layer. The evaluation metric is a combination of STOI~\cite{taal2010short} and WER, according to the official challenge rule~\cite{guizzo2022l3das22}:
\begin{equation}
     \setlength{\arraycolsep}{0.3pt}
     \text{Metric} = (\text{STOI} + (1-\text{WER}))/2
\end{equation}
The WER is computed based on the transcription of the estimated target signal and that of the reference signal, both decoded by a pre-trained Wav2Vec2.0 based ASR model~\cite{baevski2020wav2vec}. 

It can be seen from the results that the performance of Spatial-DCCRN$^*$ (the base version) is obviously better than MIMO Unet, resulting in 0.057 metric improvement with a smaller model and causal inference. Compared with frequency domain SOTA model EaBNet and time domain SOTA model FasNet, Spatial-DCCRN$^*$ yields 0.03 and 0.119 metric improvement respectively, which can prove that our model is competitive. Adding the MMF module and the AFE module leads to 0.007 and 0.011 metric gain respectively. Furthermore, the AFE module yields relatively better performance over the MMF module. This is because the angle information in multi-channel scenario is more essential. Combining AFE and MMF together, we achieve 0.016 metric improvement over the base version Spatial-DCCRN$^*$. Moreover, when more channels are available (4$\rightarrow$8), considerable improvements for those metrics can be achieved. Finally, when we extend Spatial-DCCRN to its non-causal version, the best metric score is obtained.

By comparing different loss functions, several observations can be made. 1) From experiment No. 1, 7, and 8, we can observe that when only the STOI Loss is adopted, the performance is bad. However, when combining it with other loss functions, better metric scores are obtained. It can be considered that the STOI Loss is an useful auxiliary loss function. 2) From experiment No. 6 and 7, 
it is found that the PHASEN loss yields better performance over SI-SNR. This is because WER and STOI are sensitive to spectrum distortion, while time domain loss is unstable for those metrics. 3) From experiment No. 7 and 9, compared with the single PHASEN Loss, when the SI-SNR loss is further employed, considerable improvement can be achieved. This shows that optimizing the model from both time and frequency domains is beneficial.

    \begin{table}[t]
    \centering
    \footnotesize
    \setlength\tabcolsep{6pt}
    \caption{Results of various models on ConferencingSpeech2021 challenge development set.} 
%
    \vspace{0.3cm}
    \begin{tabular}{lcccccc}
    \toprule 
    \# & Model   & Cau. &PESQ & STOI & E-STOI & SI-SNR \\ 
    \midrule 
    1 & Noisy   &  - & 1.515  & 0.823  & 0.690     & 4.474  \\ 
    2 & Baseline   &  \checkmark  & 1.999  & 0.888   & 0.780      & 9.159   \\ 
    3 & MIMO-Unet~\cite{ren2021causal}   & \checkmark & 2.215   & 0.908   & 0.817      & 9.287   \\ 
    4 & Spatial-DCCRN    &  \checkmark& \textbf{2.523}   & \textbf{0.923}   & \textbf{0.847}      & \textbf{10.167}   \\ 
    \bottomrule
    \label{tab:conference}
    \vspace{-1cm}
    \end{tabular}
    \end{table}

\begin{figure}[h]
    \centering
    \includegraphics[width=1\linewidth]{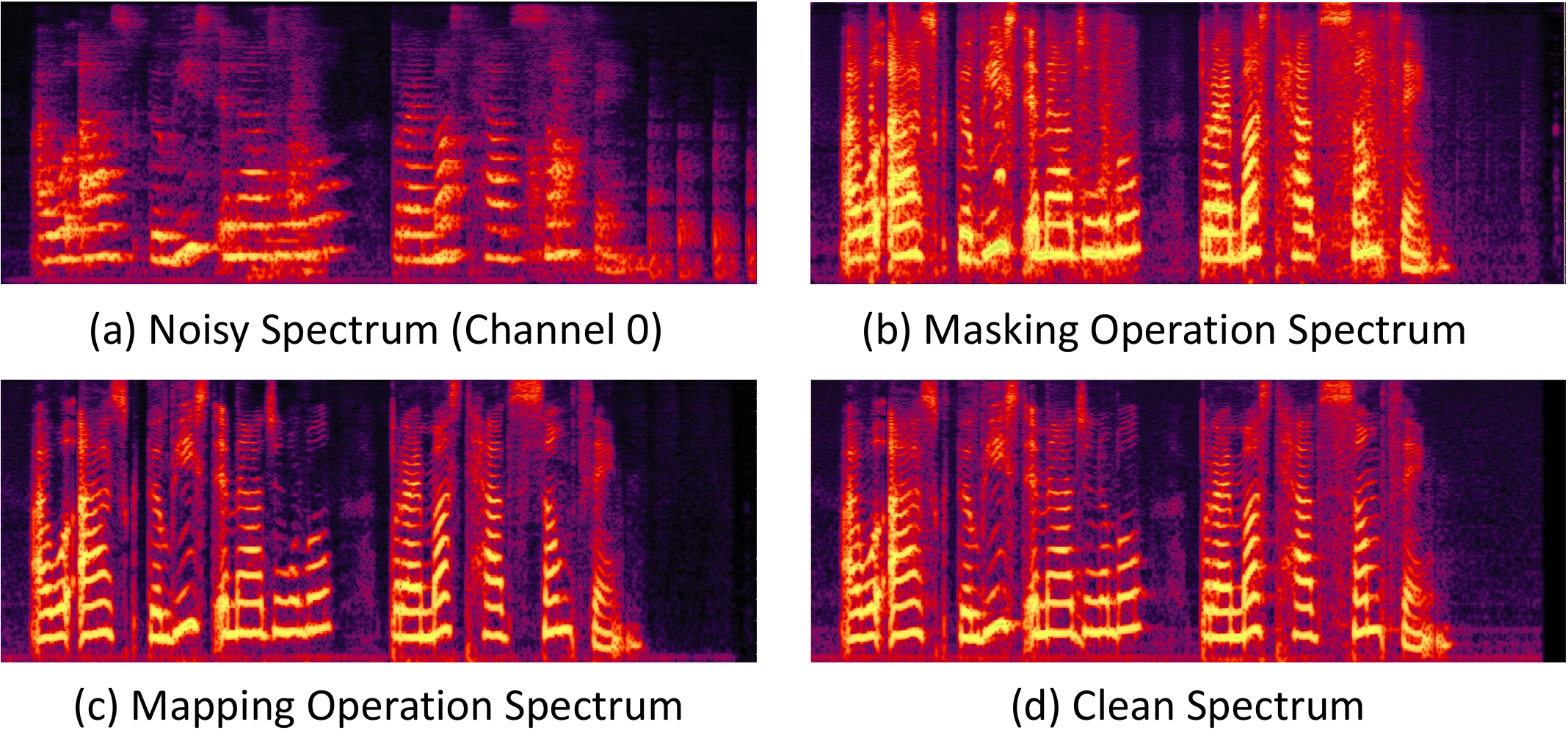}
    \caption{Masking and mapping filtering results on a testing noisy clip.}
    \label{fig:mmf}
    \end{figure}

Figure~\ref{fig:mmf} illustrates an example on the proposed masking and mapping operation (MMF).  It can be observed that the noisy spectrum was coarsely denoised and dereverbed after the masking operation. As expressed in Eq.~(\ref{form:mmf}), ${M'_{f,t}}$ is the inverse of ${c_{f,t}}$. As a result, the masking operation does good to dereverberation. It should be noted that we do not 
guide the learning of the mask operation during training. 
 As shown in Figure~\ref{fig:mmf} (c), after the masking operation, the mapping operation focuses on the lost detail from the complex-domain perspective and removes the residual noise.

We further evaluate Spatial-DCCRN on the development set provided by ConferencingSpeech2021 challenge task 1. The official baseline system of ConferencingSpeech2021 is a network composed of 3 LSTM layers and a dense layer, and the model's input is the complex spectrum of the first channel together with IPD features. Here we also take the first rank system MIMO-Unet~\cite{ren2021causal} for comparison as well measured in
PESQ~\cite{rec2005p}, STOI, E-STOI~\cite{jensen2016algorithm} and SI-SNR. 
As shown in table~\ref{tab:conference}, Spatial-DCCRN outperforms the baseline by a large margin and clearly surpasses MIMO-Unet in all metrics.
\vspace{-0.3cm}
\section{Conclusions}
In this paper, we propose a novel multi-channel complex domain denosing network -- Spatial-DCCRN, which is extended from S-DCCRN~\cite{lv2021s}. With the help of the cascaded sub-channel and full-channel processing strategy, the model can benefit from both local and global channel information processing. Importantly, an angle feature extraction module is adopted to extract frame-level angle feature, aiming at assisting the network to perceive spatial information more apparently. Finally a masking and mapping filtering method is employed to replace the traditional filter-and-sum operation. The proposed Spatial-DCCRN model obtains excellent performance with 0.956 metric score on the L3DAS22 dataset. Furthermore, Spatial-DCCRN surpasses the first-rank model MIMO-Unet on the task1 development set provided by the ConferencingSpeech 2021 challenge.~\footnote{Enhanced clips can be found from \url{https://imybo.github.io/Spatial-DCCRN/}} 

\bibliographystyle{IEEEbib}
\bibliography{strings}

\end{document}